\documentclass[letter,twocolumn]{jpsj2}

\title{Orbital-Controlled Superconductivity in $f$-Electron Systems}

\author{Katsunori \textsc{Kubo} and
Takashi \textsc{Hotta}}
\inst{Advanced Science Research Center,
Japan Atomic Energy Agency,
Tokai, Ibaraki 319-1195}

\recdate{May 16, 2006; 
  accepted June 8, 2006; published July 25, 2006}

\abst{
We propose a concept of superconductivity controlled by
orbital degree of freedom
taking Ce$M$In$_5$ ($M=$ Co, Rh, and Ir) as typical examples.
A microscopic multiorbital model for Ce$M$In$_5$
is analyzed by fluctuation exchange approximation.
Even though the Fermi-surface structure is unchanged,
the ground state is found to change significantly among
paramagnetic, antiferromagnetic, and $d$-wave superconducting phases,
depending on the dominant orbital component
in the band near the Fermi energy.
We show that our picture naturally explains the different
low-temperature properties of Ce$M$In$_5$ by carefully analyzing
the crystalline electric field states.
}

\kword{crystalline electric field, heavy-fermion superconductor,
  fluctuation exchange approximation,
  thermal expansion}

\begin{document}
\maketitle


Since the discovery of heavy-fermion superconductivity
in CeCu$_2$Si$_2$,~\cite{Steglich}
it has been one of the central issues in the research field of
condensed matter physics to unveil unconventional superconductivity
in strongly correlated electron systems.
In particular, it is important to determine a key parameter
for controlling the appearance of superconductivity.
Among heavy-fermion superconductors,
Ce$M$In$_5$ ($M=$ Co, Rh, and Ir) have potential
for systematic understanding of the mechanism and
control parameter of superconductivity,
since various ordered phases have been found
in the same crystal structure.
At ambient pressure, a relatively high superconducting transition
temperature $T_{\text{c}}=2.3$~K has been reported for
CeCoIn$_5$,~\cite{Petrovic1}
whereas $T_{\text{c}}=0.4$~K for CeIrIn$_5$.~\cite{Petrovic2}
CeRhIn$_5$ is an antiferromagnet with
a N\'{e}el temperature $T_{\text{N}}=3.8$~K at ambient pressure,
but it becomes superconducting with $T_{\text{c}} \simeq 2$~K
at pressure $p \gtrsim 1.6$~GPa.~\cite{Hegger}

The observed power-law temperature dependences
of specific heat,~\cite{Petrovic1,Movshovich,Fisher}
thermal conductivity,~\cite{Movshovich} and
spin-lattice relaxation rate \cite{Kohori1,Zheng,Kohori2,Mito}
suggest the presence of line nodes
in the superconducting gap function.
In addition, the four-fold magnetic-field angular-dependences
of thermal conductivity \cite{Izawa} and specific heat \cite{Aoki}
clearly indicate $d$-wave superconductivity for CeCoIn$_5$.
The existence of antiferromagnetism in CeRhIn$_5$ at ambient pressure
is also consistent with $d$-wave superconductivity induced
by antiferromagnetic fluctuations.

Concerning the control parameter of superconductivity,
one may first consider carrier density,
as in the case of high-$T_{\text{c}}$ cuprates.
In addition, Fermi-surface topology also plays a crucial role
in the occurrence of superconductivity
and the determination of the symmetry of the gap function.
However, the carrier densities are essentially the same among Ce$M$In$_5$,
because $f$-electron number per Ce ion is not changed by $M$.
Furthermore, Fermi surfaces observed by de Haas-van Alphen
experiments and band-structure calculations are similar
among these compounds.~\cite{Cornelius,Haga,Hall,Settai,Shishido1,Maehira}
Thus, neither the carrier density nor Fermi-surface topology
is the main control parameter of superconductivity and
antiferromagnetism in Ce$M$In$_5$.

Here, we point out another important ingredient,
the crystalline electric field (CEF) effect, in Ce$M$In$_5$.
Immediately after the discovery of Ce$M$In$_5$,
Takimoto~\textit{et al.} have stressed the importance of
the CEF effect in controlling superconductivity,~\cite{Takimoto2002}
but only level splitting was taken into account.
Note that CEF potential affects not only level
splitting, but also the CEF ground-state wave-function.
In fact, recent inelastic neutron scattering experiments have
revealed that level splittings are almost the same among
Ce$M$In$_5$, whereas CEF wave-functions are quite
different.~\cite{Christianson}
The importance of orbital states
has been discussed for Na$_x$CoO$_2$$\cdot$$y$H$_2$O,~\cite{Yanase}
but the effects of the change in orbital states
on superconductivity have not been studied systematically so far.
It is conceptually important to clarify superconductivity
controlled by orbital states.


In this Letter, we apply fluctuation exchange (FLEX) approximation
to an $f$-electron model
constructed on a square lattice for Ce$M$In$_5$.
Such perturbative theories have been applied to one-$f$-orbital
models for Ce$M$In$_5$.~\cite{Nisikawa,Ikeda,Tanaka}
In this study, we focus on the effect of a CEF using a model
including all the states with the total angular momentum $j=5/2$,
which are split into one $\Gamma_6$ and two $\Gamma_7$ doublets
under a tetragonal CEF.
Among them, the wave functions of $\Gamma_7$ states sensitively
depend on CEF potential.
We change the wave functions fixing level splittings,
and determine both $T_{\rm c}$ and $T_{\rm N}$
within FLEX approximation.
Even though the Fermi-surface structure does not change,
the ground state changes depending on the $\Gamma_7$ wave-functions
among the paramagnetic, antiferromagnetic, and $d$-wave superconducting states.
We show that the obtained phase diagram is relevant to Ce$M$In$_5$.


Using the $f$-electron basis under a cubic CEF for convenience,
we consider a three-orbital model based on a $j$-$j$ coupling scheme
on a square lattice given by
\begin{equation}
  \begin{split}
    H=&\sum_{\mib{r},\mib{\mu},\tau,\tau^{\prime},\sigma}
    t^{\mib{\mu}}_{\tau \tau^{\prime}}
    c^{\dagger}_{\mib{r} \tau \sigma}
    c_{\mib{r}+\mib{\mu} \tau^{\prime} \sigma}
    +U\sum_{\mib{r} \tau} n_{\mib{r} \tau \uparrow} n_{\mib{r} \tau \downarrow}\\
    &+(U^{\prime}/2)\sum_{\mib{r}, \tau \ne \tau^{\prime}}
    n_{\mib{r} \tau} n_{\mib{r} \tau^{\prime}}
    +H_{\text{CEF}},
  \end{split}
\end{equation}
where $c_{\mib{r} \tau \sigma}$ is the annihilation operator of
the $f$ electron at site $\mib{r}$ with
the pseudospin $\sigma$ and the orbital $\tau$,
$n_{\mib{r} \tau \sigma}=c^{\dagger}_{\mib{r} \tau \sigma} c_{\mib{r} \tau \sigma}$,
and $n_{\mib{r} \tau}=\sum_{\sigma} n_{\mib{r} \tau \sigma}$.
Note that $\sigma$ ($=\uparrow$ and $\downarrow$)
distinguishes two states in each Kramers doublet.
On the other hand, $\tau$ is introduced to distinguish
three kinds of Kramers doublets under a cubic CEF.
Here, $\alpha$ and $\beta$ denote two $\Gamma_8$,
while $\gamma$ indicates $\Gamma_7$.
We use $t^{\mib{\mu}}_{\tau \tau^{\prime}}$ for the hopping amplitude
between the $\tau^{\prime}$ state at site $\mib{r}+\mib{\mu}$
and the $\tau$ state at $\mib{r}$, where $\mib{\mu}$ is a vector
connecting nearest-neighbor sites.
We consider the hopping through $(ff\sigma)$ bonding, given by
$t^{(a,0)}_{\alpha \alpha}
=t^{(0,a)}_{\alpha \alpha}
=-\sqrt{3}t^{(a,0)}_{\alpha \beta}
=-\sqrt{3}t^{(a,0)}_{\beta \alpha}
=\sqrt{3}t^{(0,a)}_{\alpha \beta}
=\sqrt{3}t^{(0,a)}_{\beta \alpha}
=3t^{(a,0)}_{\beta \beta}
=3t^{(0,a)}_{\beta \beta}
=t$
and zero for the other cases,~\cite{Hotta}
where $t=9(ff\sigma)/28$ and $a$ is the lattice constant.
The coupling constants $U$ and $U'$ denote
intra- and inter-orbital Coulomb interactions, respectively.
Note that we ignore the Hund's rule coupling,
pair-hopping interaction, and the other interactions for simplicity.
Then, owing to the symmetry requirement, we should set $U=U'$.
We also note that $t$, $U$, and $U^{\prime}$ should be regarded
as renormalized ones for heavy electrons.~\cite{Yanase2}


The tetragonal CEF term $H_{\rm CEF}$ is given by
\begin{equation}
  \label{eq:CEF}
  H_{\text{CEF}}=\sum_{\mib{r}}
  (B^0_2 O^0_{2 \mib{r}}+B^0_4 O^0_{4 \mib{r}}+B^4_4 O^4_{4 \mib{r}}),
\end{equation}
where $B^n_m$ is the CEF parameter and $O^n_{m \mib{r}}$ is
the Stevens operator equivalent at site $\mib{r}$.~\cite{Hutchings}
We can rewrite eq.~\eqref{eq:CEF} with the basis of
the cubic CEF eigenstates as
$H_{\text{CEF}}=\sum_{\mib{r},\sigma,\tau,\tau'}
B_{\tau \tau^{\prime}}c^{\dagger}_{\mib{r} \tau \sigma}
c_{\mib{r} \tau^{\prime} \sigma}$,
where $B_{\tau \tau^{\prime}}$ is given by
an appropriate linear combination of $B^n_m$.
From the diagonalization of $H_{\rm CEF}$,
we express the CEF parameters as
$B^0_2=[-4\Delta_{67}+\Delta_7(2 \cos(2\theta)
-\sqrt{5}\sin(2\theta))]/84$,
$B^0_4=[3\Delta_{67}+\Delta_7(2 \cos(2\theta)
-\sqrt{5}\sin(2\theta))]/1260$,
and
$B^4_4=\sqrt{5}\Delta_7(\sqrt{5} \cos(2\theta)+2\sin(2\theta))/360$,
where $\Delta_{67}$ and $\Delta_7$ determine CEF energy levels,
while $\theta$ characterizes eigenstates under a CEF potential.
Then, the CEF energy levels are
$-\Delta_{67}/3-\Delta_7/2$ for the $\Gamma^{(1)}_7$ states,
$(c^{\dag}_{\mib{r} \gamma \sigma} \cos \theta+
c^{\dag}_{\mib{r} \alpha \sigma} \sin \theta)|0\rangle$,
$-\Delta_{67}/3+\Delta_7/2$ for the $\Gamma^{(2)}_7$ states,
$(-c^{\dag}_{\mib{r} \gamma \sigma} \sin \theta+
c^{\dag}_{\mib{r} \alpha \sigma} \cos \theta)|0\rangle$,
and
$2\Delta_{67}/3$ for the $\Gamma_6$ states,
$c^{\dag}_{\mib{r} \beta \sigma}|0\rangle$,
where $|0\rangle$ denotes vacuum.


Now we apply FLEX approximation, which has been extended
for multiorbital models.~\cite{Takimoto2004,Mochizuki,Mochizuki2,Yada}
Since $\sigma$ is a conserved quantity in the present model,
the Green's function does not depend on $\sigma$
and is represented by a $3 \times 3$ matrix.
In a matrix form, the Dyson-Gorkov equation is given by
$G(k)=G^{(0)}(k)+G^{(0)}(k)\Sigma(k)G(k)$,
where $G(k)$ is the Green's function and
$G^{(0)}(k)$ is the noninteracting Green's function.
Here, we have introduced the abbreviation
$k=(\mib{k},\text{i} \epsilon_n)$, where $\mib{k}$ is the momentum and
$\epsilon_n=(2n+1) \pi T$ is the fermion Matsubara frequency
with an integer $n$ and a temperature $T$.
The self-energy is given by
\begin{equation}
  \Sigma_{\tau_1 \tau_2}(k)=
  \frac{T}{N}\sum_{q \tau^{\prime}_1 \tau^{\prime}_2}
  V_{\tau_1 \tau^{\prime}_1; \tau_2 \tau^{\prime}_2}(q)
  G_{\tau^{\prime}_1 \tau^{\prime}_2}(k-q),
\end{equation}
where $N$ is the number of lattice sites,
$q=(\mib{q}, \text{i} \omega_m)$,
and $\omega_m=2m \pi T$ is the boson Matsubara frequency.
The fluctuation exchange interaction is given by
$
V(q)=
\frac{3}{2}[U^{\text{s}} \chi^{\text{s}}(q) U^{\text{s}}
-U^{\text{s}} \chi^{(0)}(q) U^{\text{s}}/2
+U^{\text{s}}]
+\frac{1}{2}[U^{\text{c}} \chi^{\text{c}}(q) U^{\text{c}}
-U^{\text{c}} \chi^{(0)}(q) U^{\text{c}}/2
-U^{\text{c}}]
$.
The matrices $U^{\text{s}}$ and $U^{\text{c}}$ are given by
$U^{\text{s}}_{\tau \tau; \tau \tau}=U^{\text{c}}_{\tau \tau; \tau \tau}=U$,
$U^{\text{s}}_{\tau \tau^{\prime}; \tau \tau^{\prime}}=-U^{\text{c}}_{\tau \tau^{\prime}; \tau \tau^{\prime}}=U^{\prime}$,
and
$U^{\text{c}}_{\tau \tau; \tau^{\prime} \tau^{\prime}}=2U^{\prime}$,
where $\tau \ne \tau^{\prime}$,
and the other matrix elements are zero.
The spin and charge parts of the susceptibility are given by
$\chi^{\text{s}}(q)=\chi^{(0)}(q)[1-U^{\text{s}} \chi^{(0)}(q)]^{-1}$
and
$\chi^{\text{c}}(q)=\chi^{(0)}(q)[1+U^{\text{c}} \chi^{(0)}(q)]^{-1}$,
respectively,
where
$\chi^{(0)}_{\tau_1 \tau_2; \tau_3 \tau_4}(q)=-\frac{T}{N}\sum_{k}G_{\tau_1 \tau_3}(k+q)G_{\tau_4 \tau_2}(k)$.


In FLEX approximation without vertex corrections,
the magnetic susceptibility is given by~\cite{Kubo}
\begin{equation}
  \chi_{\nu}(q)
  =\frac{1}{2}
  \sum_{\genfrac{}{}{0pt}{}{\tau_1 \text{--} \tau_4}{\sigma_1 \text{--} \sigma_4}}
  \chi^{\text{s}}_{\tau_1 \tau_2; \tau_3 \tau_4}(q)
  J^{\nu}_{\tau_2 \sigma_2; \tau_1 \sigma_1}
  J^{\nu}_{\tau_3 \sigma_3; \tau_4 \sigma_4}
  \sigma^{\nu}_{\sigma_1 \sigma_2}
  \sigma^{\nu}_{\sigma_4 \sigma_3},
\end{equation}
where $\nu=x$, $y$, or $z$, $\mib{\sigma}$ are the Pauli matrices,
and $J^{\nu}_{\tau_2 \sigma_2; \tau_1 \sigma_1}$ is the matrix element
of the operator of the dipole moment.
In this paper, we renormalize $J^{\nu}$ so that
the sum of squares of eigenvalues is unity, for convenience.
The linearized equation for the anomalous self-energy $\phi$
is expressed as
\begin{equation}
  \phi^{\xi}_{\tau_1 \tau_2}(k)
  =-\frac{T}{N} \sum_{k^{\prime} \tau^{\prime}_1 \tau^{\prime}_2}
  K^{\xi}_{\tau_1 \tau_2; \tau^{\prime}_1 \tau^{\prime}_2}(k,k^{\prime})
  \phi^{\xi}_{\tau^{\prime}_1 \tau^{\prime}_2}(k^{\prime}),
  \label{eq:gap}
\end{equation}
where $\xi$ denotes the pseudospin singlet (S) or triplet (T) pairing
state and the kernel is given by
$K^{\xi}_{\tau_1 \tau_2; \tau^{\prime}_1 \tau^{\prime}_2}(k,k^{\prime})
=\sum_{\tau_3 \tau_4}V^{\xi}_{\tau_1 \tau_3; \tau_4 \tau_2}(k-k^{\prime})
G_{\tau_3 \tau^{\prime}_1}(k^{\prime})
G_{\tau_4 \tau^{\prime}_2}(-k^{\prime})$.
The effective pairing interactions are given by
$
V^{\text{S}}(q)=
\frac{3}{2}[U^{\text{s}} \chi^{\text{s}}(q) U^{\text{s}}+U^{\text{s}}/2]
-\frac{1}{2}[U^{\text{c}} \chi^{\text{c}}(q) U^{\text{c}}-U^{\text{c}}/2]
$ and $
V^{\text{T}}(q)=
-\frac{1}{2}[U^{\text{s}} \chi^{\text{s}}(q) U^{\text{s}}+U^{\text{s}}/2]
-\frac{1}{2}[U^{\text{c}} \chi^{\text{c}}(q) U^{\text{c}}-U^{\text{c}}/2]
$.
The transition temperature of superconductivity is given by the temperature
where eq.~\eqref{eq:gap} has a nontrivial solution.


In the following, we show results for a $32\times32$ lattice
for $U=U^{\prime}=4t$, $\Delta_{67}=1.5t$, and $\Delta_7=t$.
The results are not so sensitive to precise values of
$\Delta_{67}$ and $\Delta_7$ as long as they are in the order of $t$.
In the calculation, we use 1024 Matsubara frequencies.
The number of $f$ electrons per site is fixed to be one,
since we are considering Ce$^{3+}$ ions.
The present model with the CEF parameter $\theta$
is invariant for $\theta \rightarrow \theta+\pi$.
In addition, when $U=U^{\prime}$,
the model is also invariant for $\theta \rightarrow -\theta$.
Thus, it is enough to consider $0 \le \theta \le \pi/2$.

\begin{figure}[t]
  \begin{center}
  \includegraphics[width=0.98\linewidth]{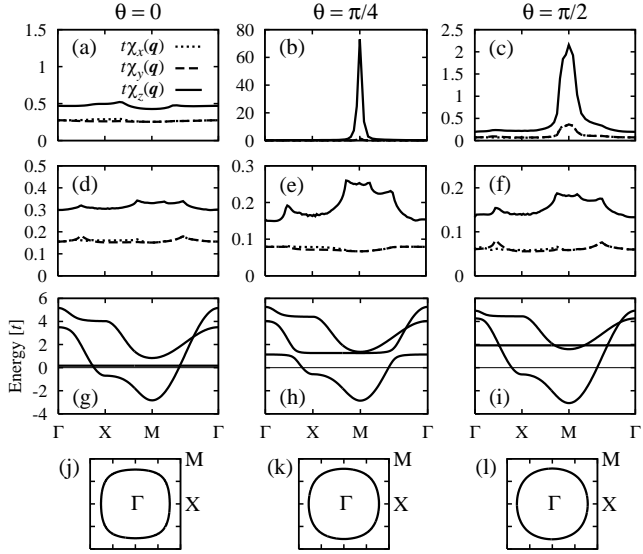}
  \caption{
    (a)--(c) Magnetic susceptibility at $U=U^{\prime}=4t$ and $T=0.02t$,
    for $\theta=0$, $\pi/4$, and $\pi/2$, respectively.
    (d)--(f) Magnetic susceptibility at $U=U^{\prime}=0$ and $T=0.02t$,
    for $\theta=0$, $\pi/4$, and $\pi/2$, respectively.
    (g)--(i) Energy band structures at $U=U^{\prime}=0$
    for $\theta=0$, $\pi/4$, and $\pi/2$, respectively.
    The Fermi energy is set at zero (thin lines).
    (j)--(l) Fermi-surface curves at $U=U^{\prime}=0$
    for $\theta=0$, $\pi/4$, and $\pi/2$, respectively.
    The parameters are set at $\Delta_{67}=1.5t$ and $\Delta_7=t$.
  }
  \label{figure:chi_chi_band_FS}
  \end{center}
\end{figure}

Figures~\ref{figure:chi_chi_band_FS}(a)--\ref{figure:chi_chi_band_FS}(c)
show the magnetic susceptibility
$\chi_{\nu}(\mib{q})=\chi_{\nu}(\mib{q},\text{i}\omega_m=0)$
for $\theta=0$, $\pi/4$, and $\pi/2$, respectively.
It is observed that $\chi_{\nu}(\mib{q})$ strongly
depends on $\theta$.
In the following, we explain this dependence from
the viewpoint of itinerant and localized orbitals.


First, let us discuss the noninteracting susceptibility $\chi^{(0)}_{\nu}(\mib{q})$
[Figs.~\ref{figure:chi_chi_band_FS}(d)--\ref{figure:chi_chi_band_FS}(f)].
In Figs.~\ref{figure:chi_chi_band_FS}(g)--\ref{figure:chi_chi_band_FS}(l),
we show noninteracting band structures and Fermi-surface curves.
Even if $\theta$ is changed,
the band that crosses the Fermi level $E_{\text{F}}$
has similar dispersion at $E_{\text{F}}$,
leading to an almost the same Fermi surface.
Note, however, that in general,
$\chi^{(0)}_{\nu}(\mib{q})$ depends on the band structure
and is not determined solely by the Fermi surface structure
(\textit{e.g.}, nesting property), particularly for multiorbital systems.
The nearly flat band composed mainly of the $\gamma$ orbital
locates near $E_{\rm F}$ for small $\theta$,~\cite{note3}
suggesting that electrons near $E_{\rm F}$ have nearly
localized character.~\cite{note1}
Thus, $\chi^{(0)}_{\nu}(\mib{q})$ for $\theta=0$
is almost flat in momentum space.
For small $\theta$, the magnitude of $\chi^{(0)}_{\nu}(\mib{q})$ is large,
since $\chi^{(0)}_{\nu} \sim \rho_0$ at sufficiently low $T$,
where $\rho_0$ is the density of states at $E_{\rm F}$,
and it becomes large when the localized character is strong.
With increasing $\theta$, the localized character is weakened.
Thus, $\chi^{(0)}_{\nu}(\mib{q})$ shows a structure in $\mib{q}$ space
and becomes smaller in magnitude.
In particular, we observe a moderate enhancement around
$\mib{q}=\mib{Q} \equiv (\pi/a,\pi/a)$
for $\theta=\pi/4$ and $\pi/2$.


Now, we consider the effect of on-site Coulomb interactions.
We expect that the Fermi surface is not markedly changed
by the Coulomb interaction.
First, note that the on-site Coulomb interactions suppress
the $\mib{q}$-independent part of $\chi^{(0)}_{\nu}(\mib{q})$
due to the correlation effect
beyond random phase approximation (RPA).
Thus, for small $\theta$,
$\chi_{\nu}(\mib{q})$ is totally suppressed
in comparison with that of RPA.
For larger $\theta$, the structure in $\chi^{(0)}_{\nu}(\mib{q})$
grows to a peak at $\mib{q}=\mib{Q}$ in $\chi_{\nu}(\mib{q})$.
On the other hand, such a peak becomes high and sharp,
when the Coulomb interactions are effectively strong
as for almost localized electrons at small $\theta$.
Thus, due to the combination of the band structure and
the effect of the Coulomb interactions,
$\chi_{\nu}(\mib{q})$ is large for a moderate value
of $\theta \simeq \pi/4$ with a peak at $\mib{q}=\mib{Q}$.

\begin{figure}[t]
  \begin{center}
  \includegraphics[width=0.92\linewidth]{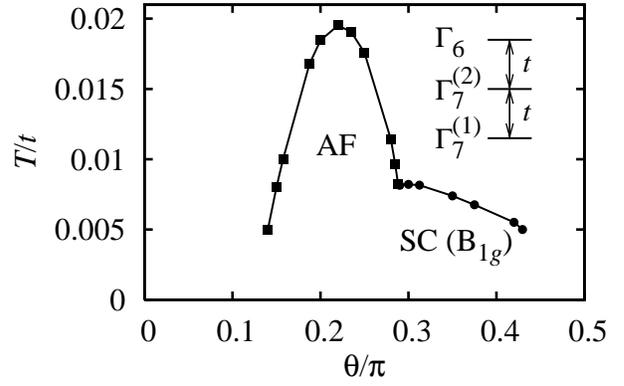}
  \caption{
    Phase diagram for
    $U=U^{\prime}=4t$,
    $\Delta_{67}=1.5t$, and
    $\Delta_7=t$.
    Solid squares denote antiferromagnetic (AF) transition temperatures,
    while solid circles denote superconducting (SC) transition temperatures
    with B$_{1g}$ symmetry.
  }
  \label{figure:PD}
  \end{center}
\end{figure}

In Fig.~\ref{figure:PD},
we show a $T$-$\theta$ phase diagram.
In this paper, we define $T_{\rm N}$ as a temperature
at which $\chi_z(\mib{Q})$ reaches $100/t$.
Note that the susceptibility does not become as large as $100/t$
for the other components $\chi_x$ and $\chi_y$,
or for $\mib{q}$ other than $\mib{Q}$.
The overall structure in Fig.~\ref{figure:PD}
is consistent with the $\theta$ dependence
of the magnetic property.
Note that a $d$-wave singlet superconducting state with B$_{1g}$
symmetry appears at $\theta \gtrsim 0.29 \pi$
next to the antiferromagnetic phase.
On the other side of the antiferromagnetic phase,
$T_{\rm c}$ is very low,~\cite{note2}
even if the superconducting phase exists.
This is consistent with the fact that
the energy scale in the small-$\theta$ region
becomes small owing to the strong localized character of electrons.

Now we discuss possible relevance of our results to Ce$M$In$_5$.
For this purpose, it is necessary to estimate $\theta$
for each material.
Among the CEF parameters, the sign of $B^4_4$ is quite important,
since the orbital state is markedly affected by the sign.
Although inelastic neutron scattering is a powerful method of
determining CEF energy levels, the sign of $B^4_4$ cannot be
determined solely by neutron scattering experiment.
Thus, we should perform CEF analysis of a quantity
sensitive to $\langle O^4_4 \rangle$,
where $\langle \cdots \rangle$ denotes the expectation value
in terms of $H_{\rm CEF}$.
Since the mode of charge distribution corresponding to finite
$\langle O^4_4 \rangle$ couples to the lattice,
thermal expansion is a good quantity for determining the sign of $B^4_4$.
Thermal expansion has been measured for Ce$M$In$_5$,
but the analysis focused only on the second-order term
$\langle O^0_2 \rangle$.~\cite{Takeuchi1,Takeuchi2,Correa,Correa2,Takeuchi3}
It is necessary to include fourth-order terms to analyze
thermal expansion, as in the case of cubic systems.~\cite{Morin}

\begin{figure}[t]
  \begin{center}
  \includegraphics[width=0.98\linewidth]{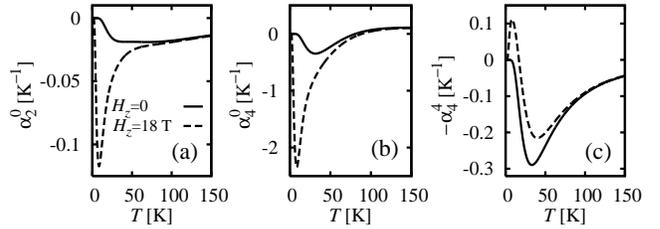}
  \caption{
    Temperature dependences of
    (a) $\alpha^0_2$,
    (b) $\alpha^0_4$, and
    (c) $-\alpha^4_4$
    for single ion under CEF potential deduced from
    neutron scattering experiment on CeRhIn$_5$.~\cite{Christianson}
    Solid curves denote the values without a magnetic field and
    dashed curves denote those in a magnetic field $H_z=18$~T
    along the $c$-axis.
  }
  \label{figure:thermal_expansion}
  \end{center}
\end{figure}

Figures~\ref{figure:thermal_expansion}(a)--\ref{figure:thermal_expansion}(c)
show the temperature dependence of
$\alpha^0_2\equiv \text{d}\langle O^0_2 \rangle/\text{d}T$,
$\alpha^0_4\equiv \text{d}\langle O^0_4 \rangle/\text{d}T$, and
$\alpha^4_4\equiv \text{d}\langle O^4_4 \rangle/\text{d}T$, respectively,
for the CEF parameters deduced from the neutron scattering experiment
for CeRhIn$_5$,~\cite{Christianson} assuming $B^4_4>0$.
A thermal expansion coefficient is given by a linear combination of
$\alpha^0_2$, $\alpha^0_4$, and $\alpha^4_4$.
For $B^4_4<0$, $\alpha^4_4$ changes its sign,
whereas $\alpha^0_2$ and $\alpha^0_4$ do not.
Although we do not show the calculated results for CeCoIn$_5$ and CeIrIn$_5$,
it has been found that the overall features are
similar among Ce$M$In$_5$.
Figures \ref{figure:thermal_expansion}(a)--\ref{figure:thermal_expansion}(c)
show that
the magnetic-field effect on $\alpha^0_2$ and $\alpha^0_4$ is strong,
but weak on $\alpha^4_4$.
From the experimental results,~\cite{Correa,Correa2}
the magnetic-field effect on the thermal expansion coefficient $\alpha_c$
along the $c$-axis for CeRhIn$_5$ is not so significant
as that observed in Figs.~\ref{figure:thermal_expansion}(a) and
\ref{figure:thermal_expansion}(b).
Rather it looks similar to that in Fig.~\ref{figure:thermal_expansion}(c).
Moreover, the minimum with a negative value in $-\alpha^4_4$
shifts toward higher temperatures under a magnetic field,
as is experimentally observed in $\alpha_c$ of CeRhIn$_5$.
Thus, we highly expect that $\alpha_c$ in Ce$M$In$_5$ is mainly
determined by $\alpha^4_4$.
For CeCoIn$_5$ and CeIrIn$_5$, $\alpha_c$ is always positive
at the temperatures reported,~\cite{Takeuchi1,Takeuchi2}
in sharp contrast to that for CeRhIn$_5$.
These observations indicate that the signs of $B^4_4$ for
CeCoIn$_5$ and CeIrIn$_5$ are the same, but are
different from that for CeRhIn$_5$.

\begin{figure}[t]
  \begin{center}
  \includegraphics[width=0.9\linewidth]{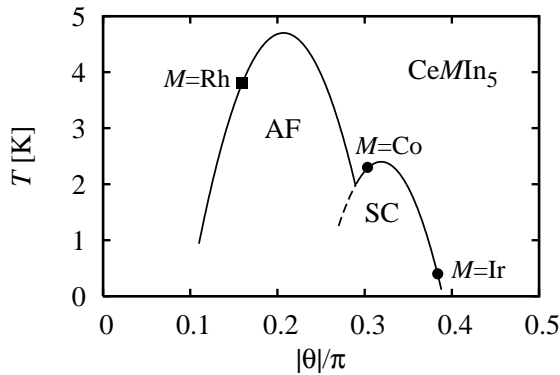}
  \caption{
    Schematic phase diagram for Ce$M$In$_5$.
    Solid square and circles denote $T_{\text{N}}$ and
    $T_{\text{c}}$, respectively.
    Curves are guides for the eyes.
    See the main text for the values of $\theta$.
  }
  \label{figure:schematic_PD}
  \end{center}
\end{figure}

For further quantitative analysis,
it is necessary to have more precise knowledge
on elastic constants and/or magnetoelastic coupling,
but it is out of the scope of this paper.
Rather, we phenomenologically assume $B^4_4>0$ for CeRhIn$_5$.
Then, $B^4_4$ is negative for CeCoIn$_5$ and CeIrIn$_5$.
By using the CEF parameters deduced from the neutron scattering
experiment,~\cite{Christianson} we obtain
$\theta=0.16\pi$ for CeRhIn$_5$,
$\theta=-0.30\pi$ for CeCoIn$_5$, and
$\theta=-0.38\pi$ for CeIrIn$_5$.
In Fig.~\ref{figure:schematic_PD},
we show a schematic phase diagram using these values of $\theta$
and experimental results for $T_{\rm c}$ and $T_{\rm N}$.
This is consistent with the theoretically determined
phase diagram in Fig.~\ref{figure:PD}.
In addition, CeRhIn$_5$ locates in a small-$\theta$ region, indicating
that $f$ electrons in CeRhIn$_5$ have a relatively localized character,
as is suggested from de Haas-van Alphen experiment.~\cite{Shishido1}
Thus, we conclude that the characteristics of Ce$M$In$_5$
can be captured by our model and that the main control parameter of
antiferromagnetism and superconductivity for Ce$M$In$_5$ is
$\theta$ which describes CEF wave-functions.
To confirm this,
it is important to experimentally determine CEF parameters
for Ce$M$In$_5$ under pressure.

In summary, we have studied the $f$-electron model
including all the states with $j=5/2$ by FLEX approximation.
We have found three kinds of ground state,
i.e., paramagnetic, antiferromagnetic,
and $d$-wave superconducting states,
by changing the CEF wave-functions characterized by the parameter $\theta$,
even if we fix level splitting.
Such phase change originating from the difference in character of
$f$-electron orbitals explains well the difference in Ce$M$In$_5$.
This finding shows that, in general,
orbital character can be a control parameter
of superconductivity, in addition to Fermi surface topology
and carrier density.
The orbital-controlled superconductivity would be realized
also in other $f$-electron materials with active orbital degrees of freedom,
and the present study will provide an important step toward
the investigation of such exotic superconductivity.


We are grateful to T. Takeuchi for sending us unpublished data
of CeCoIn$_5$.
We also thank T. Takimoto for useful discussions on FLEX calculation.
The authors are supported by a Grant-in-Aid for Scientific Research
in Priority Area ``Skutterudites''
from the Ministry of Education, Culture, Sports, Science,
and Technology of Japan.
K.~K. is also supported by the REIMEI Research Resources of
Japan Atomic Energy Agency and
by a Grant-in-Aid for Young Scientists
from the Ministry of Education, Culture, Sports, Science,
and Technology of Japan.
T.~H. is also supported by a Grant-in-Aid for Scientific Research
from the Japan Society for the Promotion of Science.



\begin{thebibliography}{99}
\bibitem{note3}
For $\theta=0$, the CEF ground state is described purely
by the $\gamma$ orbital, but a band composed of other orbitals
has lower energy at some wave-numbers due to the kinetic-energy gain.
Thus, the flat band composed of the $\gamma$ orbital locates
higher than $E_{\text{F}}$ even for $\theta=0$.
\bibitem{note1}
The hopping integrals including the $\gamma$ orbital
are zero in the present model.
They are considered to be small in the square-lattice structure
in general.
\bibitem{note2}
For instance,
we estimate $T_{\text{c}} \lesssim 0.00035t$ at $\theta=0.139 \pi$
by extrapolating the eigenvalues of the Kernel $K$ in eq.~\eqref{eq:gap}.
\end{thebibliography}
\end{document}